# Cross-Wave Profiles of Altitude and Particle Size of Noctilucent Clouds in the Case of One-Dimensional Small-Scale Gravity Wave Pattern

Oleg S. Ugolnikov

Space Research Institute, Russian Academy of Sciences, Moscow, Russia

E-mail: ougolnikov@gmail.com

**Abstract**

This paper describes the wide-field three-color observations of an expanded field of noctilucent clouds modulated by a one-dimensional gravity wave. Long wave crests were aligned by a small angle to the solar vertical in the sky. This made possible separate determination of altitude and particle size at different wave phases based on three-color photometry of noctilucent clouds. Thereby, it is possible to use simple optical imaging to record the changes in the parameters of noctilucent clouds when a short-period gravity wave passes by.

## 1. Introduction

Noctilucent clouds (NLC) – the highest and coldest clouds in the Earth's atmosphere – appear in the summer mesosphere of middle and high latitudes at the altitudes of 80–85 km. Their brightness and density usually increase closer to pole; however, clouds cannot be seen in the sky in the polar midnight sun and bright nights. NLC are better to observe at the latitudes of 50–60°, where the summer twilights are dark.

To be seen in the sky, the cloud must contain ice particles with a radius of no less than 20–30 nm. Their formation requires a temperature of below 145–150 K, which is common near the summer pole, but can be less probable in mid-latitudes. NLC were not observed until 1885; this can be the effect of the long-term cooling trend in the mesosphere (Thomas, Olivero, 2001). While the temperature in the mid-latitude upper mesosphere is usually above the frost point on average, it is characterized by variations driven by different wave mechanisms: gravity waves with periods from several minutes to 10 hours (Jensen and Thomas, 1994), tidal semi-diurnal and diurnal waves (Pertsev *et al.*, 2015), 5-day planetary waves (Gadsden, 1985; Merkel *et al.*, 2003), and 27-day solar cycle (Robert *et al.*, 2010). Ice particles are formed in regions with minimal temperatures. The basic characteristics: the total water content, layer height, particle concentration, and mean radius – change with the wave phase, which leads to brightness variations and a wavelike structure often observed in NLC.

If the gravity wave has a short period of about 1 hour or less and a wavelength of not more than several hundred kilometres, one can observe several maxima in the sky during one twilight period. In this case, one can trace the evolution of the cloud in time during the wave period or spatially over one or several wavelengths. A short gravity wave can decrease the mean NLC albedo since the particle sublimation in the temperature maxima is faster than its growth in the cold phase of the wave (Jensen and Thomas, 1994; Rapp *et al.*, 2002; Gao *et al.*, 2018).

When NLC cover most of the sky hemisphere, wide-angle photo imaging can be used to determine cloud properties – the mean radius and altitude. The particle size can be estimated by wide-angle polarization analysis (Ugolnikov *et al.*, 2016; Ugolnikov and Maslov, 2019). Another approach is three-band (RGB) photometry using an all-sky color camera (Ugolnikov *et al.*, 2017). This method was developed in (Ugolnikov, 2021), the basic principles were confirmed by radiative transfer analysis. In those papers, the altitude and particle radius were determined for the NLC field in the sky as a whole. It is also worth doing it for individual elements of the NLC field. It can be done if the field is easily separated into different segments with different properties, and each of them must cover a wide range of scattering angles. The optimal configuration is the 1-D (parallel) wave pattern. The present paper contains an analysis of such NLC ensemble observed in 2022.

**2. Observations**

Bright NLC with a fine 1-D wave structure were observed in central Russia (55.6°N, 36.6°E) in the morning of July 15, 2022. Clouds appeared near the local midnight when the sky became clear. Coming out from the shadow of the Earth, NLC were expanded to the most part of the sky reaching the scattering angles of 160° during the twilight. The sky hemisphere field was recorded by three identical Xiaomi Mi Sphere cameras with a field of view of about 190° and a resolution of 3456x3456 pixels. The angular resolution was about 19 pixels per degree; the data were binned in 1° circles. Effective wavelengths of B, G, and R wide spectral channels of the camera for the case of NLC were equal to 474, 529, and 587 nm, respectively. Detailed information on the spectral bands can be found in (Ugolnikov, 2021). For better result accuracy, combined data of all three cameras were used. Exposure times decreased from 4 seconds in the night (solar zenith angle more than 99°) to 0.25 seconds in the light stage of twilight (solar zenith angle 94–95°) when NLC were disappearing on the bright sky background.

Star images on the night and dark twilight frames were used to fix the position parameters of each camera and local atmospheric transparency; the procedure is analogous to (Ugolnikov, 2021). The maximal sky point zenith angle in the analysis was increased up to 70°, except for certain cases due to landscape restrictions. Figure 1 shows an example of the sky image, taken by one of three cameras, at the moment 0hUT (solar zenith angle 97.3°). One can see that the major part of the sky is covered by bright NLC that have a remarkable ray-like structure. These rays are in fact almost parallel lines in space, and the observed picture is the display of a one-dimensional gravity wave pattern.

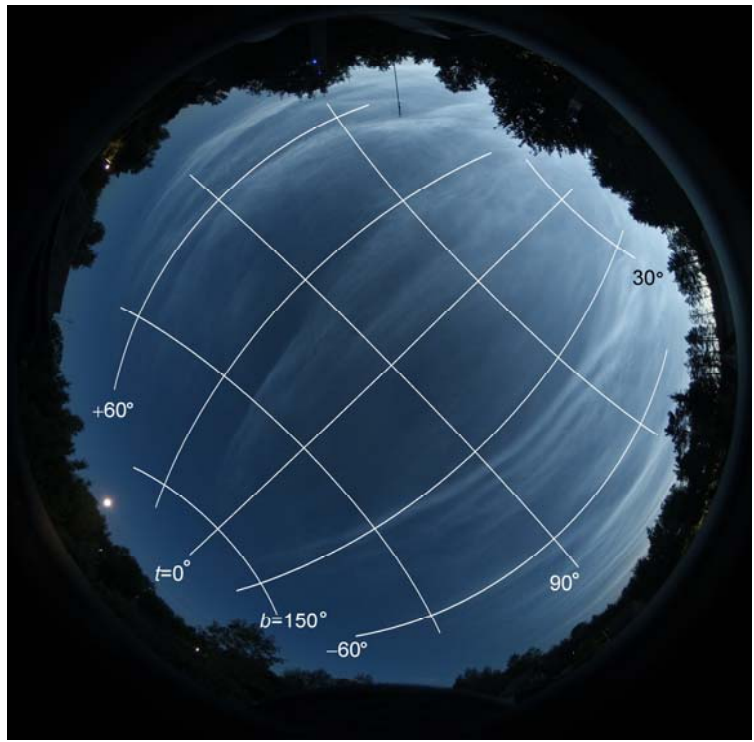

*Figure 1. Example of sky image of NLC (July, 15, 2022, 0h UT) with the denotation of sky coordinate system used in this work.*

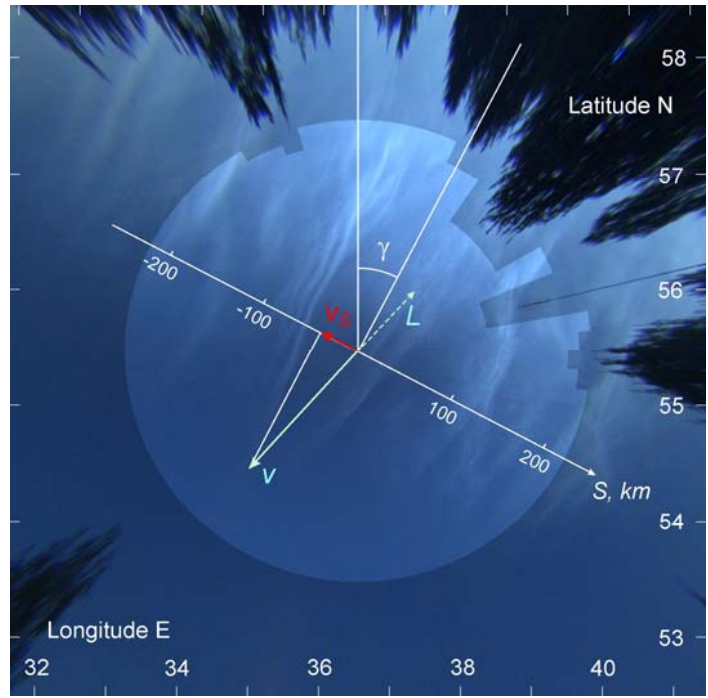

*Figure 2. Map of noctilucent clouds above the observation site with denotation of pattern velocity (July, 15, 0h UT).*

The procedure of estimating altitude and particle size based on NLC color data is described and applied to the whole registered NLC field (Ugolnikov, 2021). The results show the average values of altitude and effective particle size. The median particle radius can also be determined, if we assume that the particle size distribution is lognormal with the width parameter of σ=1.4 (von Savigny and Burrows, 2007); this is about 0.5 of the effective particle radius. The altitude measurements are based on NLC color changes as the cloud immerses into or comes out of the shadow of low atmospheric layers. The effect is defined by Rayleigh and aerosol scattering, $O_3$ and $NO_2$ absorption and can be modelled using EOS Aura/MLS data (pressure/temperature and ozone profiles, NASA GES DISC, 2022), Suomi NPP/OMPS data (aerosol, the same reference), and typical $NO_2$ profiles for the given date, time, and latitude (Gruzdev and Elokhov, 2021). Effect of color change can be analysed globally (whole sky hemisphere) or locally, focusing on the limited sky area if it contains bright NLC fragments. However, particle radius determination requires NLC color measurements in a wide range of scattering angles θ; the dependency of the NLC color on cosθ is built (Ugolnikov *et al*., 2017; Ugolnikov, 2021). The final particle radius value is the average by the cloud field being analysed.

In the case described in this paper, long wave crests were fortunately inclined by the small angle (5° on average) to the solar vertical, and each of them covered a wide range of θ. This makes it possible to run the procedure for each phase of the wave separately and to see how the altitude and particle size change as the wave passes by. However, one must know the wave velocity and choose the optimal coordinate system for this analysis. It will be done in the following chapter of the paper.

**3. Coordinate system and wave motion**

Figure 2 shows the map of the NLC field calculated from the image in Figure 1. The average altitude of NLC obtained by global color analysis (Ugolnikov, 2021), 82.5 km, is taken for this map. The analysis field (zenith angle up to 70° with landscape restrictions) is marked bright. One can see that the bright NLC arcs visible in Figure 1 are nearly straight lines almost parallel to each other. Based on the correlation analysis of the field, one can find the average angle between the wave pattern and meridian: γ=27.7°. Since the azimuth of the Sun during the NLC observation period changed from –157° to –140°, the angle between the pattern and solar direction did not exceed 12°, vanishing in a definite moment during the observations.

Ugolnikov (2021) used the sky coordinate system with the pole coinciding with the Sun, so the scattering angle θ was one of two angular coordinates. For this work, the most convenient choice is the system with the pole at the horizon at the azimuth equal to (–180°+γ). This point is the apex of NLC arcs in Figure 1. The sky point is characterized by the angles $b$ and $t$, where the module of $b$ is the angular distance between the apex and the sky point.

One of the most complicated problems in NLC photometry is the separation of the cloud field against the twilight sky background. The procedure was described in detail in (Ugolnikov *et al.*, 2021), where it was applied to the analysis of polar stratospheric clouds, and used for NLC (Ugolnikov, 2021) in the same form. In those papers, the scans of the sky along the lines θ=const are separated into background and cloud ranges by statistical iterations, and then the background is approximated by a 6-degree polynomial and subtracted.

In this paper, scans of the sky brightness are made along the lines $b$=const (equal angular distance from the apex of the pattern). Since the apex is close to the Sun in the sky, these lines are inclined a little to the lines θ=const. Another coordinate, $t$, is almost constant in the wave crest at any definite time. The lines $b$=const are perpendicular to the waves; this provides the maximal brightness variation and the most exact separation. The procedure was carried out for each $b$ value with a step interval of 1°, the resolution by $t$ is also 1°.

Figure 3 shows the profile of sky brightness along the line $b$=90° for the image shown in Figure 1. The approximation of the background is shown by the lines. The data on background-subtracted NLC brightness $I_{BGR}(b, t)$ is binned in 5° arcs by $b$ for each $t$, the criterion for data selection is $I_{BGR}(b, t) > 0$ for all five $b$ profiles, and the total NLC to background ratio of the arc is not less than 0.02. For each data point, its time $UT$, the local solar zenith angle (visible from the cloud, assuming its average altitude 82.5 km) $z_L$, and the scattering angle θ are recorded. The data of all three cameras are processed together.

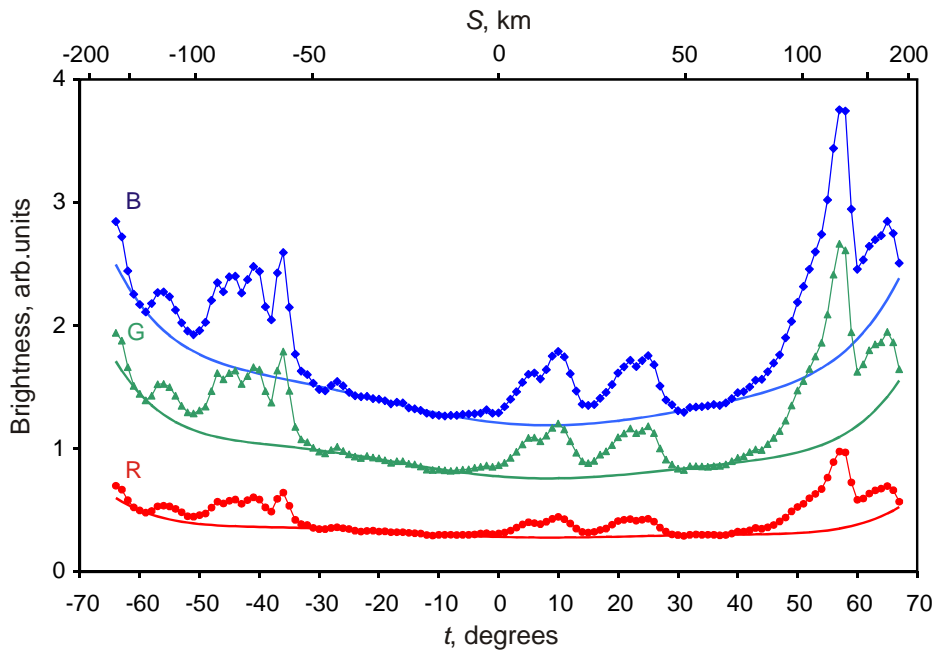

*Figure 3. Example of sky brightness profiles (0h UT, b=90°) with approximation of sky background. B band brightness is multiplied by 1.5 for convenience.*

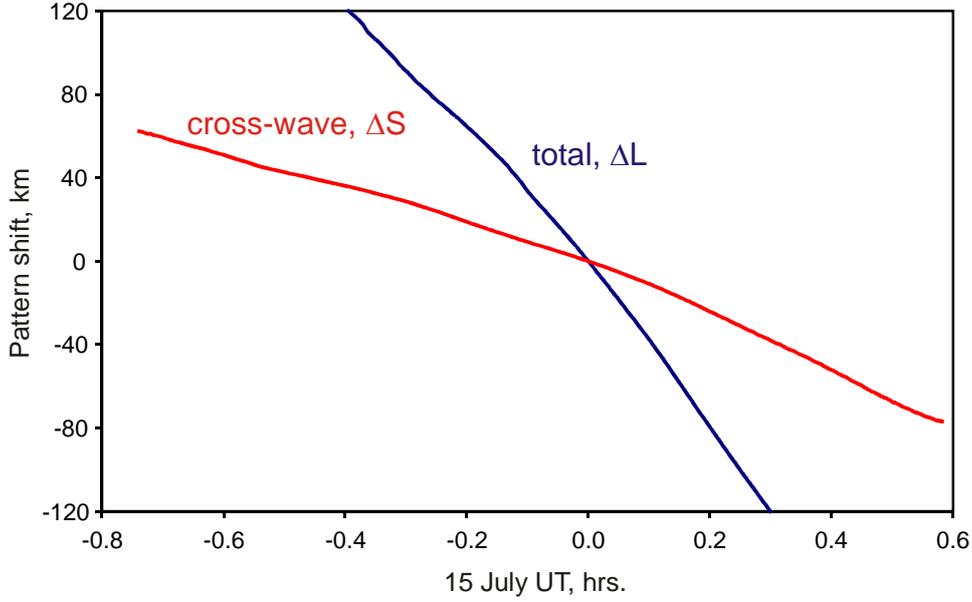

*Figure 4. Wave pattern motion during the observations.*

The sky coordinate system described above is fixed; it does not move with the Sun or the wave pattern of NLC. To separate the data with different wave phases, one must take into account NLC motion. It is done by a correlation analysis of different NLC images with one-pixel resolution. Total phase velocity *v* is found to be around 100 m/s, which increases during the observation period. The direction of wave pattern motion is SW with a mean position angle of 223° clockwise from the north (see Figure 2). Such value and direction of visible NLC velocity are typical for clouds in northern mid-latitudes (Dalin *et al.*, 2022). The dependency of the wave pattern shift $\Delta l$ on time is shown in Figure 4 (blue line). For the purposes of this study, one must take note of the cross-wave velocity component $v_S$, which is determined with better accuracy than the total or wave-aligned component of velocity. The dependency of the pattern cross-wave shift $\Delta S$ on time is also shown in Figure 4.

If the element of the wave pattern has the cross-wave coordinate $S_0$ at the moment UT=0, it will have the coordinate $S = S_0 + \Delta S (\Delta T)$ at the moment UT+$\Delta$T. The line $S$=const will be seen in the sky as the arc with a fixed coordinate *t*:

$$t(S_0, UT) = \arctan \frac{(R+H)\sin\lambda}{(R+H)\cos\lambda - R} \approx \arctan \frac{S}{H - \dfrac{S^2}{2(R+H)}}.$$

Here *R* is the Earth's radius (assumption of the spherical Earth is enough for the required accuracy of *t*, about 1°), *H* is the average altitude of NLC, $\lambda = S/(R+H)$. The approximation in the right part of the equation is suitable for the whole observable sky area. For zenith angles less than 60°, an even more simple "flat" relation $t=\arctan(S/H)$ can be used. *S* values are also denoted in Figures 2 and 3 (both figures correspond to the moment UT=0, and $S=S_0$ in this case). Analysing the data with the spatial phase coordinate $S_0$, the measurements are taken at the coordinate *t* depending on the image time UT. It is the base for altitude and effective particle size profiles depending on $S_0$. The results are discussed in the following chapter.

## 4. Results

Figures 5 and 6 show the profiles of "photometric altitude" and effective particle size in the direction perpendicular to the wave pattern as defined by color analysis G-B and R-B, respectively. The profile of total brightness of all NLC recorded during the observations at a given phase coordinate $S_0$ is also shown in both figures. The character maxima of this profile are also seen in Figure 2. A good agreement of results for two color pairs is observed. The altitude of clouds does not vary so much; however, it can be seen decreasing in the wave crests, where NLC brightness is maximal. It is expected since the NLC layer can become wider and reach lower altitudes around 80 km with high water vapour concentration near the temperature minima.

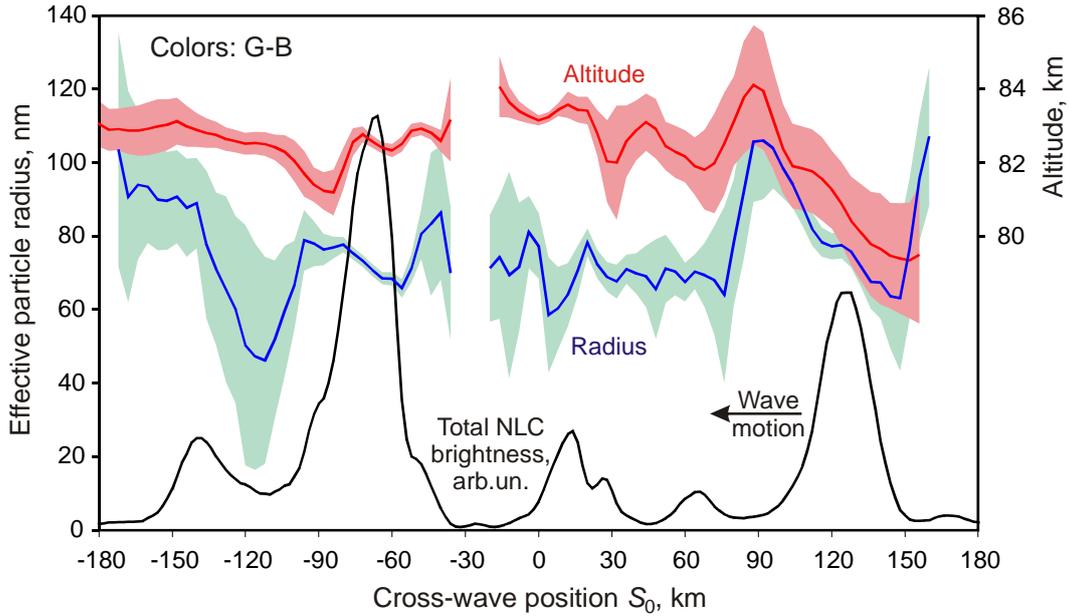

*Figure 5. Cross-wave profiles of NLC altitude and effective particle size by analysis of color ratio G-B.*

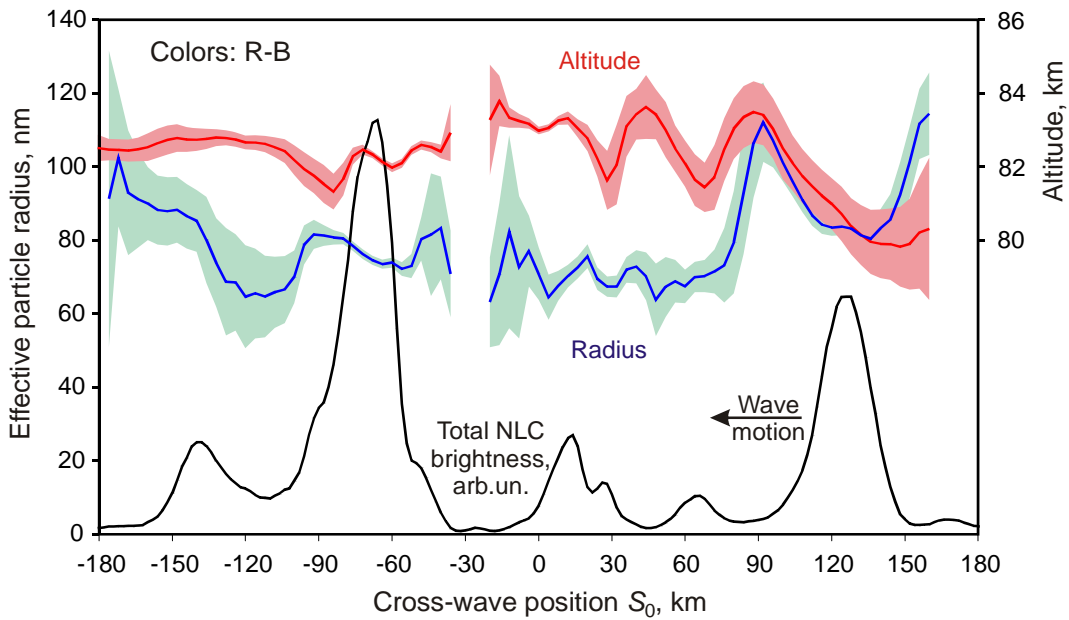

*Figure 6. Cross-wave profiles of NLC altitude and effective particle size by analysis of color ratio R-B.*

The behaviour of the effective particle size is even more noteworthy. In the most part of the $S_0$ range, the mean radius is close to the value obtained by the general all-sky method (Ugolnikov, 2021) for the same twilight, 75 nm (the mean radius for lognormal distribution with σ=1.4 is 38 nm), especially in the wave crests. The particle radius does not significantly vary in different wave maxima, confirming the general approach in (Ugolnikov, 2021). However, the effective radius can change between the crests, in the range of minimal NLC brightness. Most remarkable is its increase; this can seem surprising.

Typical wavelength of the pattern is about 80 km, as we can see in Figures 5 and 6. Knowing the cross-wave velocity (27 m/s on average), one can find the typical period of gravity wave: 50 min. It is a short period, and the difference in the times of growth and evaporation of small and large particles can play a significant role. A decrease in temperature leads to the growth of subvisible particles to a larger size (Gao *et al.*, 2018), leading to an increase in brightness. As the wave passes by, tiny particles sublimate faster than large particles, and the effective size can increase over time. A similar effect is seen in CARMA model data even for longer periods (Merkel *et al.*, 2009). Lidar analysis of a number of particle size distributions (Baumgarten *et al.*, 2010) shows that the cases with large mean radii are characterized by the constant or even decreasing width of Gaussian distribution or fast decreasing width σ, if the lognormal distribution is considered. This can take place in the case of loss of small particles while temperature increases.

AIM/CIPS data on a 20-km wave (Gao *et al.*, 2018) even show a remarkable anti-correlation of the particle size with NLC albedo. One can note that the two brightest waves shown in Figures 5 and 6 are followed by a local increase in the effective particle radius. However, one must notice that in the case of low NLC brightness, the radius determination is not so accurate.

## 5. Discussion and conclusion

In this paper, the procedure for determining the particle size and altitude of NLC based on all-sky RGB imaging was used independently in different parts of the NLC field. This opportunity was given by a remarkable parallel pattern of NLC waves on the morning of July 15, 2022. The pattern was oriented by a small angle to the solar vertical, and a number of measurements were made in a wide range of scattering angles for each phase of the wave. This configuration was especially important for the procedure.

The wave pattern velocity was revealed by image correlation analysis; the value and direction were found to be typical for NLC. The angle between the pattern and velocity was also small (about 20°), and it was another lucky factor: the pattern shift during the observation period was significantly less than the linear size of the observational NLC field, and all basic structures were permanently observed during the twilight. It is important for the altitude measurements, where the basic observational effect is the change in the NLC color during the twilight, as the cloud immerses into (or leaves) the shadow of lower atmospheric layers.

The results help to see the evolution of cloud characteristics as the wave front passes it by. The most interesting effect is the possible increase in the effective particle size beyond the wave; this can be related to different times of creation and sublimation of small and large particles. However, bright parts of clouds have similar sizes and altitudes. This confirms the applicability of the general method suggested by Ugolnikov (2021), where the NLC parameters are defined as average over the observed part of the sky. Such observations are cost-effective and simple; they can be the basis of a systematic survey of NLC and trends of their parameters, that is one of most actual problem of NLC study now (Berger and Lübken, 2015). The data can be supplied by NLC triangulation, if a number of remote cameras are used. This can be also noteworthy if a fine wave pattern is observed, and one can explore each wave element separately.